\documentclass[tikz,letterpaper,english,reprint,aps,pra,superscriptaddress,notitlepage,onecolumn,nofootinbib]{revtex4-1}

\usepackage[T1]{fontenc}
\usepackage[latin9]{inputenc}
\usepackage{hyperref}
\usepackage{color}
\usepackage{verbatim}
\usepackage{amsmath,braket}
\usepackage{amsthm,bm}
\usepackage{bbm}
\usepackage[normalem]{ulem}
\usepackage{amssymb}
\usepackage{graphicx}
\usepackage{subfigure}
\usepackage{subcaption}

\newtheorem{Pro}{Proposition}
\newtheorem{theorem}{Theorem}

\makeatletter
\def\<{\langle}
\def\>{\rangle}

\pdfpageheight\paperheight
\pdfpagewidth\paperwidth

\usepackage{graphicx}

\usepackage{pgfplots}
\usepgfplotslibrary{fillbetween}

\usepackage{tikz}
\usetikzlibrary{patterns}
\usepackage{pgfplots}
\usetikzlibrary{positioning}
\usetikzlibrary{decorations.markings}
\usetikzlibrary{arrows,decorations.text}
\usetikzlibrary{backgrounds,fit,decorations.pathreplacing}

\usepackage{babel}

\newcommand{\ketbra}[1]{\ket{#1}\!\bra{#1}}

\makeatother

\usepackage{babel}
\usetikzlibrary{positioning, arrows.meta, calc}
\begin{document}
	\title{Optimality of generalized Choi maps in $M_3$}
 	\author{Giovanni Scala}
\email{giovanni.scala@ug.edu.pl}
\affiliation{Dipartimento Interateneo di Fisica, Politecnico di Bari, 70126 Bari, Italy}
\affiliation{INFN, Sezione di Bari, 70126 Bari, Italy}
\affiliation{International Centre for Theory of Quantum Technologies, University
of Gdansk, Jana Ba\.zy\'nskiego 1A, Gda\'nsk, 80-309, Poland}
\author{Anindita Bera}
	\affiliation{Center for Theoretical Physics, Polish Academy of Sciences, Aleja Lotnik\'{o}w 32/46, 02-668 Warsaw, Poland
}
	
\author{Gniewomir Sarbicki}
	\affiliation{Institute of Physics, Faculty of Physics, Astronomy and Informatics,
		Nicolaus Copernicus University, Grudziadzka 5/7, 87-100 Toru\'{n},
		Poland}
	\author{Dariusz Chru\'{s}ci\'{n}ski}
	\affiliation{Institute of Physics, Faculty of Physics, Astronomy and Informatics,
		Nicolaus Copernicus University, Grudziadzka 5/7, 87-100 Toru\'{n},
		Poland}
	\date{\today}
	\begin{abstract}
		\noindent 
  
A family of linear positive maps in the algebra of $3 \times 3$ complex matrices proposed recently in \href{https://arxiv.org/abs/2212.03807}{Bera {\em et al.} arXiv:2212.03807} is further analyzed. It provides a generalization of a seminal Choi nondecomposable extremal map in $M_3$. We investigate when generalized Choi maps are optimal, i.e. cannot be represented as a sum of positive and completely positive maps. This property is weaker than extremality, however, it turns out that it plays a key role in detecting quantum entanglement.

	\end{abstract}
	\maketitle
	
	\section{Introduction}
 
Positive maps play an important role both in physics and mathematics providing
generalizations of $*$-homomorphisms, Jordan homomorphisms, and conditional expectations. Moreover, it
serves as a powerful tool for characterizing quantum entanglement and hence plays a key role in
various aspects of quantum information theory. Although there exist several results regarding the classification of positive maps \cite{Stormer1,Stormer2,Bhatia,Paulsen,Evans,Tom1,Tom2,Tom3,majewski01,kossakowski07,kossakowski09,KOREA,kye12,sarbicki14,rutowski17,majewski20,bera22}, the complete analysis of
positive maps still remains an open question. In this direction, examples
of positive linear maps given by Choi \cite{Choi1,Choi2,Choi3} play a prominent role in
investigating the structure of the positive cones of positive linear maps.
In this paper, we consider a generalization of the family of linear positive maps in $M_3$  proposed by Cho {\em et al.} \cite{Korea1}  which are generalizations of the Choi maps.  The positivity of such maps is rigorously discussed in Ref.~\cite{bera22}. Our main focus is to find the condition for optimality.


Le us first discuss the properties/known results regarding the positive maps.  
 Let $M_n$ denote a matrix algebra of $n \times n$ matrices over the complex field $\mathbb{C}$. A linear map $\Phi: M_n \to M_m$ is said to be positive if $\Phi(X) \geq 0,~ \forall~ X \geq 0$.  Positive maps from $M_n$ to $M_m$  form a convex cone ${\cal P}_{n,m}$, but the structure of this cone is very intricate and still not fully understood.  A positive map $\Phi \in {\cal P}_{n,m}$ is called $k$-positive if the extended map 
\begin{equation}
\Phi^{(k)} := {\rm id}_k \otimes \Phi : M_k(M_n) \to M_k(M_m) ,    
\end{equation}
is positive. Here, ${\rm id}_k$ denotes an identity map in $M_k$. A map is completely positive if it is $k$-positive for all $k=1,2,\ldots$. It was proved by Choi \cite{Choi-75} that  $\Phi$ is complete positive if $\Phi^{(k)}$ is positive for $k = {\rm min}\{n,m\}$. A positive map is said to be decomposable \cite{Stormer1}  if it  can be  decomposed as 
$    \Phi = \Phi_1 + \Phi_2\circ {\rm T}$,
where $\Phi_1,\Phi_2$ are completely positive, and ${\rm T}$ denotes a matrix transposition. Due to Woronowicz \cite{Woronowicz}, it is well-known that the cones ${\cal P}_{2,2},~{\cal P}_{2,3},~{\cal P}_{3,2} $  consist of decomposable map.  Choi provided the  first example of a non-decomposable map \cite{Choi1,Choi2,Choi3} in $\mathcal{P}_{3,3}$  

\begin{equation}
\label{Choi1}
  \Phi(X) = \begin{bmatrix}
   x_{11} + x_{22} &-x_{12} & -x_{13} \\ -x_{21} & x_{22}+x_{33} & - x_{23} \\ -x_{31} & -x_{32} & x_{33}+x_{11}
\end{bmatrix},
\end{equation}
with $X =(x_{ij}) \in M_3$.  Later on, the above map (\ref{Choi1}) was  generalized to a 3-parameter family of maps \cite{Korea1}

\begin{equation}   \label{Korea}
\Phi_C(X) = D_C(X) - X  ,    
\end{equation}
where $D_C(X)$ is a diagonal matrix with $D_{ii}= \sum_{j=1}^3 C_{ij} x_{jj}$, and $C=\{C_{ij}\}$ is  $3\times 3$ circulant matrix 
\begin{equation}
\label{abc}
C = \begin{bmatrix}
 a&b&c    \\ c&a&b \\ b&c&a
\end{bmatrix} , \ \ \ a,b,c \geq 0 .
\end{equation}
The map (\ref{Korea}) will reduce to (\ref{Choi1}) for the parameter values $a=2, ~b=1,$ and $c=0$.
One has the following \cite{Korea1}

\begin{theorem}
\label{TH-Korea}
The map $\Phi_C$ defined by (\ref{Korea})  is positive if and only if  the
following three conditions are satisfied:
\begin{equation}
    a \geq 1,\qquad a + b + c \geq 3, \qquad \mbox{if}~ a \leq 2, ~\mbox{then}~~ bc \geq (2-a)^2.
\end{equation}
\end{theorem}
It is well known that the Choi map is extremal in ${\cal P}_{3,3}$  \cite{HaLAA}. 
In this paper, we analyze the optimality property of a class of positive maps in $\mathcal{P}_{3,3}$, which we discuss in the next section. A map $\Phi \in \mathcal{P}_{n,m}$
is called optimal if for any completely positive map $\Psi$, the map $\Phi - \Psi$ is no longer positive and extremal if the only positive maps  $\Phi - \Psi$ such that $\Psi$ is positive are of the form $\alpha \Phi$ with $\alpha \in [0,1]$. It is important to mention here that optimality is less restrictive than extremality. Any extremal map is optimal but the converse needs not be true  \cite{KOREA,TOPICAL}. For example, the well-known reduction map \cite{reduction1,reduction2}
   $ R_n(X) = \mathbf{I}_n {\rm Tr}\,X - X$ in $\mathcal{P}_{n,n}$ is optimal for all $n\geq 2$ \cite{KOREA,TOPICAL} but it is extremal for $n=2$ only. In  Ref. \cite{Lew} the following sufficient condition for optimality was provided
   
\begin{theorem}
\label{TH1} 
Let $\Phi \in \mathcal{P}_{n,m}$. Consider a family of product vectors $x_i \otimes y_i \in \mathbb{C}^n \otimes \mathbb{C}^m$ such that
\begin{equation}\label{xy=0}
  \< y_i, \Phi( \overline{x}_i \overline{x}_i^\dagger) y_i \> = 0 .
\end{equation}
If  $\mathrm{span}\{x_i \otimes y_i\}=\mathbb{C}^n \otimes \mathbb{C}^m$ then $\Phi$ is optimal. 
\end{theorem}
Maps possessing a full spanning set $\{x_i \otimes y_i\}$ satisfying (\ref{xy=0}) are said to have {\it spanning property} \cite{Lew}. Note that the spanning property is only a sufficient condition for optimality, but not a necessary one.  There exist optimal maps without spanning property \cite{S71,S72,Remik,PRA-2022,LAA23, BERA23}. For example,  the Choi map is optimal without having the spanning property  \cite{S71,S72}.
In this paper, we analyze the optimality property of a large class of positive maps in $\mathcal{P}_{3,3}$ which provides a generalization of the Choi map of Eq. \eqref{Choi1}.  Moreover,  we investigate optimality by checking which maps possess the spanning property. 
 

\section{The class of positive maps}
\label{positive_map}

Following Eq.~(\ref{Korea}) the class of map that we consider in our work is defined as follows:
\begin{equation}   \label{general}
\Phi_W(X) = D_W(X) - X,    
\end{equation}
where  $D_W(X)$ is a diagonal matrix with the diagonal entries  $\sum_{j=1}^3 w_{ij} x_{jj}$, and $W = \{w_{ij}\}$ is  $3\times 3$ matrix such that $w_{ij} \geq 0$
and
\begin{equation}  \label{ww}
    \sum_{i=1}^3 w_{ij} = \sum_{j=1}^3 w_{ij} = w ,
\end{equation}
i.e. $W/w$ is a doubly stochastic matrix. Clearly when $W=\{w_{ij}\}$ is circulant, then this class of maps reduces to the form in Eq.~(\ref{Korea}). 

To show the map $\Phi_W$ is positive, it is enough to check the positivity for rank-one positive matrices $X=\psi\psi^\dagger$:
	\begin{equation} \label{pos1}
	    \forall \psi \in \mathbb{C}^3, \quad \Phi_W(\psi\psi^\dagger) = \mathrm{diag} \left( \sum_{i=1}^3 w_{1i} |\psi|^2_i, \sum_{i=1}^3 w_{2i} |\psi|^2_i, \sum_{i=1}^3 w_{3i} |\psi|^2_i \right) -\psi\psi^\dagger \ge 0.
	\end{equation}
	Let us denote $x_i = |\psi_i|^2$, $z_i = \sum_{j=1}^3 w_{ij} x_j$.
	The principal minor $M_I$ of $\Phi_W(\psi\psi^\dagger)$, constructed from the rows and columns of indices from the set $I \subset \{1,2,3\}$ is equal to
	\begin{equation}
	\label{mix}
        M_I (\vec x) = \prod_{i \in I} z_i - \sum_{i \in I} x_i \prod_{j \in I \setminus \{ i \}} z_j,
        \qquad\mbox{with  } \prod_{j \in \emptyset} z_j=1.
	\end{equation}
	Observe that, the positivity depends only on modules of $x_i$, and not on their phases. Moreover, the norm of $\psi$, equals to $x_1 + x_2 + x_3$, is irrelevant for positivity. Hence one can rewrite the condition (\ref{pos1}) as:
	\begin{equation} \label{minors}
	    \forall \vec x \in \Delta^2,~ \ \forall I \subset \{1,2,3\},~ \ M_I(x) \ge 0, 
	\end{equation}
	where $\Delta^2$ denotes the 2-simplex with vertices $x_i$.  Below we provide the conditions under which the map $\Phi_W$ is positive (for details cf. Ref.~\cite{bera22}).

\begin{theorem} \label{th1} Let $W$ be a $3\times 3$ matrix with non-negative elements satisfying (\ref{ww}) together with the following Hessian  condition
\begin{equation}  \label{www}
 \frac 12 \left( w - \sqrt{ ({\rm Tr}\, W - 2 w)^2 + 3\delta^2} \right)^2 \geq 
 \sum_{i=1}^3 (w_{ii}-w_{i+1,i+1})^2,~~\delta := |w_{ij} - w_{ji}|  ,  \ \ \ i\neq j .   
\end{equation}
Then  $\Phi_W$ is positive if and only if 
\begin{itemize}
    \item $w_{ii} \geq 1$ for $i=1,2,3$ (vertex conditions)
    \item  $\sqrt{(w_{ii}-1)(w_{jj}-1)} - \sqrt{w_{ij}w_{ji}} \geq 1$, for $i\neq j$  (edge conditions)
    \item $w\geq 3$ (interior condition).
\end{itemize}
	  
\end{theorem}
Again, if $w_{ij}=C_{ij}$  is cirulant, then condition (\ref{www}) is trivially satisfied, and   Theorem \ref{th1} reduces to Theorem \ref{TH-Korea}.

Now,  followed by Ref.~\cite{bera22}, we consider the new parameterization of the matrix $W$. As $W/w$ is a doubly stochastic matrix   the  Birkhoff theorem implies 
\begin{equation}
    W = 
    \begin{bmatrix}
 a & b & c    \\ c & a & b \\ b & c & a
\end{bmatrix} 
+ d 
\begin{bmatrix}
 0 & 1 & 0    \\ 1 & 0 & 0 \\ 0 & 0 & 1
\end{bmatrix} 
+ e 
\begin{bmatrix}
 0 & 0 & 1    \\ 0 & 1 & 0 \\ 1 & 0 & 0
\end{bmatrix}
+ f 
\begin{bmatrix}
 1 & 0 & 0    \\ 0 & 0 & 1 \\ 0 & 1 & 0 
\end{bmatrix}  
\nonumber 
 = \left(\begin{array}{ccc}
		a+f & b+d & c+e\\
		c+d & a+e & b+f\\
		b+e & c+f & a+d
	\end{array}\right) ,
\end{equation}
with  $\{a,b,c,d,e,f\}$ satisfying $a+b+c+d+e+f=w$.
Note that the parameters $\{d,e,f\}$ perturb the circulant part $\{C_{ij}\}$. However, this representation is not unique. Actually, performing the following {\em gauge} transformation
\begin{equation}
    \{a,b,c\} \to \{a+\xi,b+\xi,c+\xi\} \ , \ \ \ \{d,e,f\} \to \{d-\xi,e-\xi,f-\xi\} ,
\end{equation}
one does not affect the matrix elements $w_{ij}$'s (which have to be non-negative). Clearly, the normalization condition $a+b+c+d+e+f=w$ is gauge invariant. Here, we fix the gauge condition
\begin{equation} \label{gauge}
    d+e+f=0 .
\end{equation}
Note that in such a gauge, we have ${\rm Tr}\, W = 3a$, $w = a+b+c$, and $\delta = |b-c|$. 

Under this parametrization,  the Hessian condition (\ref{www}) can be rewritten as follows
\begin{equation} \label{hess2}	         
	        \frac 16 \left( a+b+c - \sqrt{(a-2b-2c)^{2}+3(b-c)^{2}} \right)^2 \ge d^2 + e^2 + f^2.
	    \end{equation}
The vertex condition is $w_{ii}\geq 1$ and $w_{ij} \geq 0$ for $i \neq j$ if

\begin{equation} \label{vert}
	        d,e,f \ge \max\{1-a,-b,-c\} = -\min \{a-1, b, c \} \stackrel{df}= -\mu .
	    \end{equation}
	 Edge conditions have the following form
\begin{align}
	        \sqrt{(a+f-1)(a+e-1)} + \sqrt{(b+d)(c+d)} \ge 1, 
	        \label{edge1} \\
	       \sqrt{(a+d-1)(a+f-1)} + \sqrt{(b+e)(c+e)} \ge 1,
	       \label{edge2} \\
	        \sqrt{(a+e-1)(a+d-1)} + \sqrt{(b+f)(c+f)} \ge 1, 
	        \label{edge3}
	    \end{align}
and the interior one
	    \begin{equation}
	        a + b + c \ge 3. \label{interior}
        \end{equation}

\begin{figure}[h!]
        \centering
        \begin{minipage}{0.48\textwidth}
        \includegraphics[width=.8\textwidth]{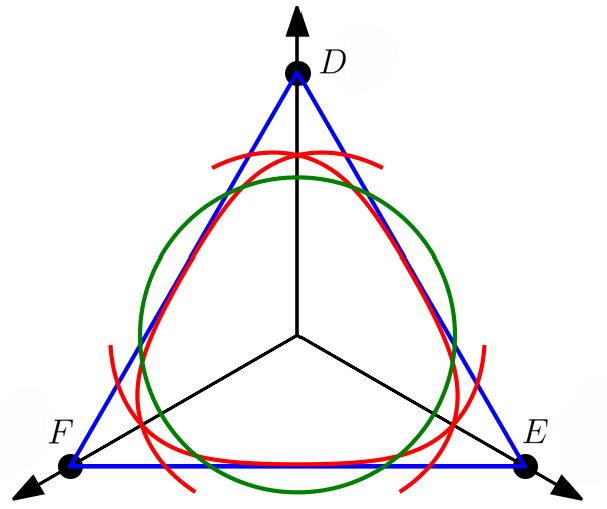} 
        \caption*{(a) frontal axonometric projection}
        \end{minipage}
        \begin{minipage}{0.48\textwidth}
        \includegraphics[width=.7\textwidth]{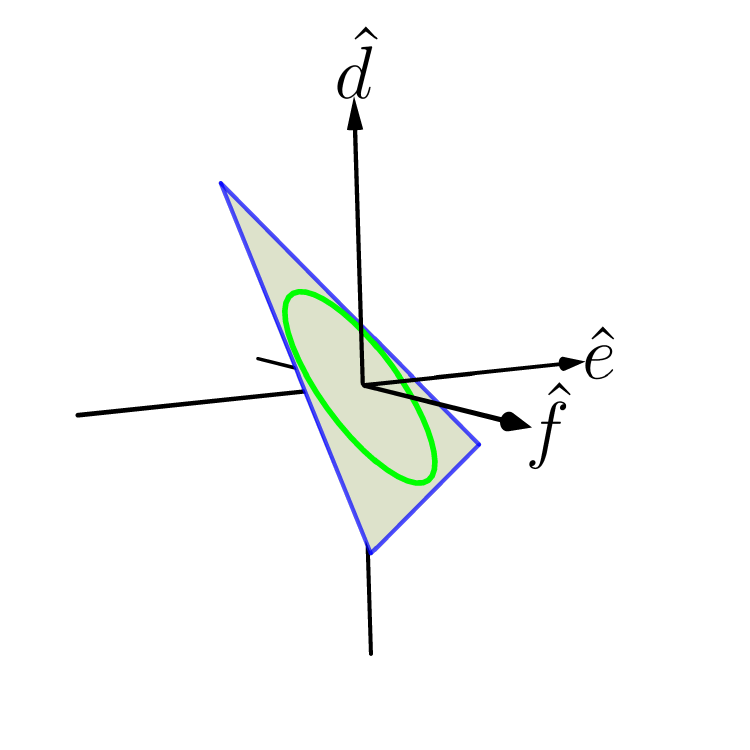} 
        \caption*{(b) knight axonometric projection}
        \end{minipage}
        \caption{ 
        A typical shape of the set of admissible points in the $d+e+f=0$ plane is an intersection of three configurations: a circle (green), 
        {Alice} (red), and 
        {Bob} (blue). Points outside the intersection of the triangles refer to non-positive maps. The positivity of maps referring to points in the intersection of the triangles, but outside the circle are not known. Here the values of the parameters are $a=1.7$, $b=0.9$, $c=0.5$ for the LHS figure. The RHS shows the plane $d+e+f=0$ from a non-frontal viewpoint. }
        \label{fig:triangles1}
    \end{figure}

    It turns out that the  condition \eqref{hess2} describes a circle of  radius
    \begin{equation}
        r_H = \frac 1{\sqrt{6}}\left( a+b+c - \sqrt{(a-2b-2c)^{2}+3(b-c)^{2}} \right)
    \end{equation}
    on the plane $d+e+f=0$, centered in the origin as shown in Fig. \ref{fig:triangles1}. 
    Note also that, if $a \ge 1, b,c \ge 0$ then the condition (\ref{vert}) cuts off an equilateral triangle on the plane $d+e+f = 0$.
    {This triangle will be further referred as Bob.}
    Its vertices are $(2\mu,-\mu,-\mu)$, $(-\mu,2\mu,-\mu)$, $(-\mu,-\mu,2\mu)$ and we denote them $D$, $E$, $F$ respectively.
    
      The inequality (\ref{edge1}) describes a shape 
    symmetric w.r.t. the height of the triangle passing through its vertex $D$. Similarly, the inequalities (\ref{edge2}) and (\ref{edge3}) describe identical shapes, but symmetric w.r.t. the heights passing through vertices $E$ and $F$ respectively.
    The intersection of these three shapes is then 
    a curved equilateral triangle, whose sides are arcs defined by saturations of the inequalities \eqref{edge1}-\eqref{edge3}. 
    {This triangle will be further referred as Alice.}

    {Alice} is non-empty iff the point $d=e=f=0$ satisfies \eqref{edge1} - \eqref{edge3}, which yields:
    \begin{equation}\label{eq:edge_kye}
        2-a \le \sqrt{bc}.
    \end{equation}
    In this way, we recover Theorem \ref{TH-Korea} for the map $\Phi_C$ ($d=e=f=0$).
    
    In the space of parameters $\{d,e,f\}$ of Fig. \ref{fig:triangles1}, the typical shape of the set of admissible points in the plane $d+e+f=0$ is an intersection of the circle (hessian condition (\ref{hess2})), 
    {Alice and Bob}, mentioned in green, red and blue colors, respectively.
    
	
 \section{Optimality of the map}
	\label{optimality}
	
Let us now examine the optimality property in the considered family of maps by checking the spanning condition (cf.  Theorem \ref{TH1}). For that purpose, we are looking for vectors $\ket{\phi^*}\otimes\ket{\psi}$ such that $\Phi_W(\ketbra{\psi})\ket{\phi} = 0$. Indeed, we will find the spanning vectors observing that the equation $\det\Phi_W(\ketbra{\psi}) = 0$ has non-zero solutions, i.e. $\ket{\psi}\neq0$ iff one of the seven conditions (\ref{minors}) saturate, which translates into saturations of three vertex conditions (\ref{vert}), three edge conditions (\ref{edge1}-\ref{edge3}) or single interior condition (\ref{interior}). 
  

\subsection{Saturations of interior, edge, and vertex conditions}

	


\begin{itemize}

\item {\bf Interior condition:} the interior condition (\ref{interior}) is saturated for $\vec x = (x_1,x_2,x_3)$ with $x_i=x_j$. Hence if $\psi = [\exp(i\alpha_1), \exp(i\alpha_2), \exp(i\alpha_3)]$, for arbitrary phases $\alpha_1, \alpha_2, \alpha_3$, then $\mbox{ker} \Phi_W (\ket{\psi}\bra{\psi})$ is spanned by $\phi = \psi$ {(See appendix \ref{sec:interior})}. Vectors $\ket{\psi^*} \otimes \ket{\psi}$ span the 7-dimensional subspace orthogonal to the following subspace

$$ {\rm span}\{e_1 \otimes e_1 - e_2 \otimes e_2 , \  e_2 \otimes e_2 - e_3 \otimes e_3\} . $$

\item {\bf Edge condition:} The edge condition is derived under the assumption that exactly one $x_i = 0$ in Eq.~(\ref{mix}). Let us assume that: $x_1 = 0, x_2,x_3 \ne 0$. It leads to the edge condition: $\sqrt{(a-1
+e)(a-1+d)}+\sqrt{(b+f)(c+f)} \ge 1$ and its saturation implies (up to a multiplicative constant):
\begin{equation}
    x_1 = 0, \qquad x_2 = ((c+f)(a+e-1))^{-\frac 12}, \qquad x_3 = ((b+f)(a+d-1))^{-\frac 12}
\end{equation}
and hence

\begin{equation}
  \ket  \psi = \left[ 0,\sqrt{x_2} \exp(i\alpha_2), \sqrt{x_3} \exp(i\alpha_3) \right]
  = \left[0,\frac{\exp(i\alpha_2)}{\sqrt[4]{(c+f)(a-1+e)}}, \frac{\exp(i\alpha_3)}{\sqrt[4]{(b+f)(a-1+d)}} \right],
\end{equation}
for arbitrary phases $\alpha_2,\alpha_3$. Now  $\Phi_W(\ket{\psi}\bra{\psi})$ takes the form:
\begin{equation}
	\left[ \begin{array}{ccc} (b+d)x_2 + (c+e)x_3 & 0 & 0 \\ 0 & (a-1+e)x_2 + (b+f)x_3 & -\sqrt{x_2x_3}\, \mathrm{e}^{\mathrm{i}(\alpha_2-\alpha_3)} \\ 0 & -\sqrt{x_2x_3}\, \mathrm{e}^{-\mathrm{i}(\alpha_2-\alpha_3)} & (c+f)x_2 + (a-1+d)x_3 \end{array} \right]
\end{equation}
and its kernel is spanned by the vector $\ket{\phi}$: 
\begin{equation}\label{eq:edgepsi}
	\ket{\phi} = [0, \sqrt{x_2x_3}\mathrm{e}^{\mathrm{i}\alpha_2}, \left((a-1+e)x_2+(b+f)x_3\right)\mathrm{e}^{\mathrm{i}\alpha_3}] \sim [0, \frac{\mathrm{e}^{\mathrm{i}\alpha_2}}{\sqrt[4]{(b+f)(a-1+e)}}, \frac{\mathrm{e}^{\mathrm{i}\alpha_3}}{\sqrt[4]{(c+f)(a-1+d)}}].
\end{equation}
{After multiplying by $\sqrt[4]{(b+f)(c+f)(a+d-1)(a+e-1)}$} one has
\begin{equation}\label{eq:edgespan}
	\ket{\phi^*} \otimes \ket{\psi}
	= \left[0,0,0,0,\sqrt[4]{\frac{a-1+d}{a-1+e}}, \sqrt[4]{\frac{c+f}{b+f}}\mathrm{e}^{\mathrm{i}(\alpha_3-\alpha_2)}, 0, \sqrt[4]{\frac{b+f}{c+f}}\mathrm{e}^{\mathrm{i}(\alpha_2-\alpha_3)}, \sqrt[4]{\frac{a-1+e}{a-1+d}}\right].
\end{equation}

These vectors span the following 3-dimensional subspace 
\begin{equation}\label{subspace_edge}
	\mathrm{span}\{e_2,e_3\}\otimes\mathrm{span}\{e_2,e_3\} \cap \left[ \sqrt[4]{\frac{a-1+e}{a-1+d}} e_2\otimes e_2 - \sqrt[4]{\frac{a-1+d}{a-1+e}} e_3\otimes e_3\right]^\perp.
\end{equation}
See appendix \ref{sec:edgecondition} for more details.

\item {\bf  Vertex condition:} the vertex conditions concern vectors having two vanishing components. If a vertex condition is saturated, say $w_{33} = 1$, then the $\Phi_W(\ket{\psi}\bra{\psi})$ is singular for $\ket\psi = e_3$ and takes the form:
 \begin{equation}
    \left[\begin{array}{ccc} c+e & 0 & 0 \\ 0 & b+f & 0 \\ 0 & 0 & a+d-1 \end{array}\right]= \left[\begin{array}{ccc} c+e & 0 & 0 \\ 0 & b+f & 0 \\ 0 & 0 & 0 \end{array}\right] ,
 \end{equation}
 due to $a+d-1 = w_{33}-1=0$.  Its kernel is spanned by $\ket\phi = e_3$ and the vectors $|\phi^* \rangle\otimes\ket\psi$ span the 1-dimensional subspace $\mathrm{span}\{e_3 \otimes e_3\}$. See appendix \ref{sec:vertexcondition} for more details.

\end{itemize}

\subsection{Combinations of saturations providing spanning}
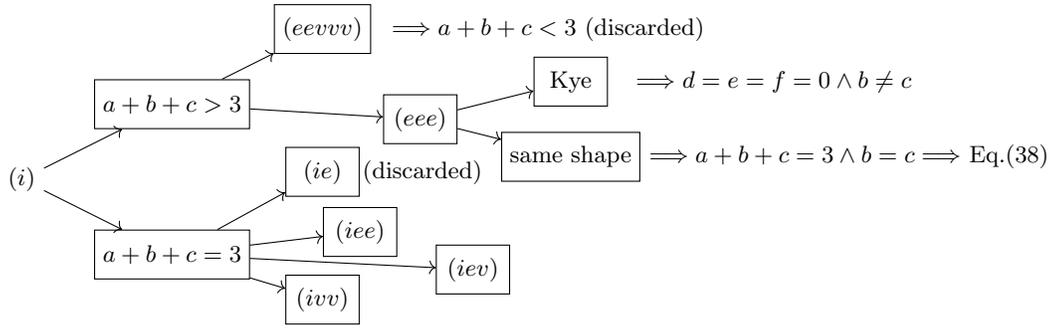
\begin{figure}[h]
    \centering
\begin{tikzpicture}[node distance=1cm and 1.5cm, auto]
\tikzset{box/.style={rectangle, draw, text centered, minimum height=2em, minimum width=3em}}

 \node (start) {\((i)\)};
 
 \node[box] (upperbox) at ($(start)+(2,1)$) {\(a+b+c>3\)};
 \node[box] (lowerbox) at ($(start)+(2,-1)$) {\(a+b+c=3\)};

 \node[box] (eevvv) at ($(upperbox)+(2,1)$) {\((eevvv)\) };
    \node (violation) at ($(eevvv)+(3,0)$) {\(\Longrightarrow a+b+c<3\)~(discarded)};

 \node[box] (eee) at ($(upperbox)+(3.3,-0.2)$) {\((eee)\)};
     \node[box] (kye) at ($(eee)+(2,0.5)$) {Kye};
    \node (Kyecond) at ($(kye)+(2.7,0)$) {\(\Longrightarrow d=e=f=0\land b\neq c\)};
    \node[box] (sameshape) at ($(eee)+(2,-0.5)$) {same shape};
    \node (sameshapecond) at ($(sameshape)+(3.7,0)$) {\(\Longrightarrow a+b+c=3\land b= c \Longrightarrow \)  Eq.\eqref{eq:edge_eq}};
 \node[box] (ie) at ($(lowerbox)+(2,1.1)$) {\((ie)\)};
 \node (iecond) at ($(ie)+(1.33,0)$) {(discarded)};
 \node[box] (iee) at ($(lowerbox)+(2.5,0.3)$) {\((iee)\)};
 \node[box] (iev) at ($(lowerbox)+(4,-0.2)$) {\((iev)\)};
 \node[box] (ivv) at ($(lowerbox)+(2,-0.6)$) {\((ivv)\)};

 \draw[->] (start) -- (upperbox);
 \draw[->] (start) -- (lowerbox);

 \draw[->] (upperbox) -- (eevvv);
 \draw[->] (upperbox) -- (eee);
    \draw[->] (eee) -- (kye);
    \draw[->] (eee) -- (sameshape);

\draw[->] (lowerbox) -- (ie);
\draw[->] (lowerbox) -- (iee);
\draw[->] (lowerbox) -- (iev);
\draw[->] (lowerbox) -- (ivv);
 
\end{tikzpicture}
    \caption{{No saturation of the interior condition $(i)$ implies the study of spanning property for the saturation of $(eevvv)$ and $(eee)$ to span 9-dimensional space. The former violates the interior condition, thus it must be discarded. The latter gives \textit{(eee) Kye case}, i. e. Eq. \eqref{eq:edge_kye} restoring the family of theorem \ref{TH-Korea} and \textit{(eee) same shape case} discussed in Fig. \ref{fig:3edge_optimal}. When the $(i)$ condition is saturated then condition $(ie)$ is discarded since (e) contributes with one additional spanning vector and the cases to study for the spanning property are $(iee)$, $(iev)$, and $(ivv)$ depicted in Fig. \ref{fig:lin}.}}
    \label{fig:combination}
\end{figure}

Among the seven conditions discussed in the previous section, if two or more above conditions are saturated, then the spanning subspace is an algebraic sum of the corresponding subspaces mentioned above. In what follows we denote the interior condition by (i), the edge condition by (e), and the vertex condition by (v). In Fig. \ref{fig:combination} we represent all the cases that we study to satisfy the spanning property. Assume first, that the interior condition (i) is not saturated.   Then to obtain a 9-dimensional space we have to collect three edge saturations (eee), or two edges and three vertex saturations (eevvv). The simultaneous saturation of three vertex conditions implies $0 = d = e = f = - \min \{ a-1, b, c \} \Rightarrow a=1 \ \land \ b=c=0$, the interior condition is violated hence we are left with only one nontrivial scenario, that is, the saturation of (eee). 

\begin{itemize}

\item {\bf Spanning by the saturation of three edge conditions $(eee)$:}  The three edge conditions can be simultaneously saturated if they intersect at the same point ($d=e=f=0$ by the symmetry
) or if the three equations describe the same curve. 

{\it The (eee) Kye case:} Here we consider the scenario when the three equations (\ref{edge1}-\ref{edge3}) become the same and give the family \cite{Korea1} provided in Eq.~(\ref{Korea}) parameterized by circulant matrices. The three edge conditions reduce to a single one: $a+\sqrt{bc}=2$ and from Eq. \eqref{subspace_edge} the spanning space reads
\begin{align}
	\mathrm{span} \left\{
	\left[0,0,0,0,1,\sqrt{\frac cb}e^{i\alpha_{32}},0,\sqrt{\frac bc}e^{-i\alpha_{32}},1\right] 
	\right\}
	&
	\oplus
	\mathrm{span} \left\{
	\left[1,0,\sqrt{\frac bc}e^{i\alpha_{13}},0,0,0,\sqrt{\frac cb}e^{-i\alpha_{13}},0,1\right]
	\right\}
	\nonumber \\
	&
	\oplus
	\mathrm{span} \left\{
	\left[1,\sqrt{\frac cb}e^{i\alpha_{21}},0,\sqrt{\frac bc}e^{-i\alpha_{21}},1,0,0,0,0\right]
	\right\}
	,
\end{align}
is $9$-dimensional if $b,c \ne 0$. Here $\alpha_{ij}=\alpha_i-\alpha_j$. Figure \ref{fig_kye} shows the shape of the boundary of the family proposed in \cite{Korea1}. The green region characterizes the parameters of the maps that saturate the edge condition. While it is a surface of the cone centered in $[a,b,c] = [3,0,0]$, any such map can be decomposed as a combination of a map on the ellipse: $a+b+c=3 \ \land \ a+\sqrt{bc}=2$ and a map on the hyperbola: $a=1, bc=1$ \cite{KOREA}.

    \begin{figure}[h!]
    \begin{tikzpicture}
    \node at (0,0) {\includegraphics[width=.75\textwidth]{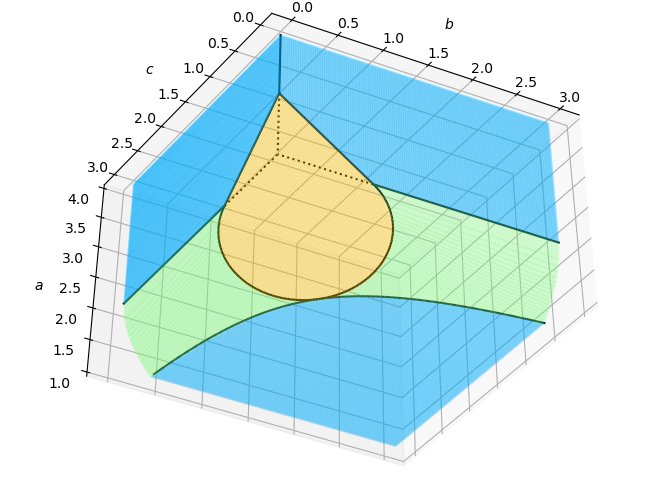}};
    \node[align=center] at (-3,1.3) {vertex \\ $b=0$};
    \node[align=center] at (2,2) {vertex \\ $c=0$};
    \node[align=center] at (.5,-2) {vertex \\ $a=1$};
    \node[align=center] at (-.5,.5) {interior \\ $a+b+c=3$};
    \node[align=center] at (2.5,-.5) {edge \\ $a+\sqrt{bc}=2$};
    \node[align=center] at (-2.5,-1) {edge \\ $a+\sqrt{bc}=2$};
\end{tikzpicture}
    \caption{
    The boundary of the family \cite{Korea1} ($d=e=f=0$)  provided in Eq.~(\ref{Korea}).  The blue parts are the maps saturating one of the vertex conditions. The green part consists of maps saturating the edge condition and the orange region consists of maps saturating the interior condition (see details in \cite{code}).
    }
    \label{fig_kye}
    \end{figure}


{\it The (eee) same shape case:} Now we consider  the scenario where the three Eqs (\ref{edge1}-\ref{edge3}) represent the same curve on the plane defined by \(d+e+f=0\). {This implies that the gradients of the edge functions w.r.t. the variables \(d, e, f\), when projected onto the 2-dimensional subspace \(\{ d,e,f: d+e+f=0\}\), must match to each other.} Thus, let us define the following three functions

\begin{align}
	       F_1 := \sqrt{(a+f-1)(a+e-1)} + \sqrt{(b+d)(c+d)} , 
	        \label{edgeF1} \\
	     F_2:=  \sqrt{(a+d-1)(a+f-1)} + \sqrt{(b+e)(c+e)} ,
	       \label{edgeF2} \\
	      F_3:=  \sqrt{(a+e-1)(a+d-1)} + \sqrt{(b+f)(c+f)}. 
	        \label{edgeF3}
\end{align}
such that the three edge conditions read $F_k \geq 1$ for $k=1,2,3$. Let us calculate their gradients w.r.t. variables $d,e,f$ projected to the subspace $e+d+f=0$, that is,
$$ \vec \nabla F := \left. (\partial_d F,\partial_e F,\partial_f F)\right|_{d+e+f=0} .$$
One finds {(see appendix \ref{gradient_expression} for the calculations):}
\begin{align}\label{gradients}
    \left. 
    \vec \nabla F_1 \right|_{F_1=1}
    = \left( a+b+c-1-\frac{b+c+2d}{\sqrt{(b+d)(c+d)}} \right) 
    \left[ \begin{array}{ccc} -2 \\ 1 \\ 1 \end{array} \right] 
    - 3 \left[ \begin{array}{ccc} d \\ e \\ f \end{array} \right],
    \\
    \left. 
    \vec \nabla F_2 \right|_{F_2=1}
    = \left( a+b+c-1-\frac{b+c+2e}{\sqrt{(b+e)(c+e)}} \right) 
    \left[ \begin{array}{ccc} 1 \\ -2 \\ 1 \end{array} \right] 
    - 3 \left[ \begin{array}{ccc} d \\ e \\ f \end{array} \right],
    \\
    \left. 
    \vec \nabla F_3 \right|_{F_3=1}
    = \left( a+b+c-1-\frac{b+c+2f}{\sqrt{(b+f)(c+f)}} \right) 
    \left[ \begin{array}{ccc} 1 \\ 1 \\ -2 \end{array} \right] 
    - 3 \left[ \begin{array}{ccc} d \\ e \\ f \end{array} \right].
\end{align}
These vector fields are identically equal iff the terms in the round brackets are zeros:
\begin{equation}\label{equal_vec}
    a+b+c-1
    = \frac{b+c+2d}{\sqrt{(b+d)(c+d)}}
    = \frac{b+c+2e}{\sqrt{(b+e)(c+e)}}
    = \frac{b+c+2f}{\sqrt{(b+f)(c+f)}},
\end{equation}
which implies $b=c, ~a+b+c=3$. In this case, formulas (\ref{edge1}-\ref{edge3}) reduce to a single formula (see appendix \ref{derivation}):
\begin{equation}\label{eq:edge_eq}
    6(b-1)^2=\frac 32 (a-1)^2 \ge d^2+e^2+f^2.
\end{equation}
{and Alice becomes a circle.} It implies the  hessian condition of Eq. \eqref{hess2} that reads
 \begin{align}
        6b^2\ge d^2+e^2+f^2,  & ~~ \text{if} ~~ b\le \frac 12, \label{hessian1}\\
        6(b-1)^2 \ge d^2+e^2+f^2,  & ~~ \text{if} ~~ b> \frac 12.
        \label{hessian2}
    \end{align}

{
On the plane $d+e+f=0$ the inequality \eqref{eq:edge_eq} is saturated on the circle of the radius $\sqrt{3/2} (a-1)$. If $a=1$ the theorem \ref{TH-Korea} is obtained. If $a \in (1,5/3)$ then this circle, e.g. at the bottom of Fig. \ref{fig:3edge_optimal} (see also Fig. \ref{fig:3edge_optimal}.i, belongs to the boundary of the positive maps set and each of its points corresponds to an optimal map in (eee) scenario. Indeed, it gives 3 spanning vectors from the saturation of each edge condition. If $a=5/3\simeq 1.66$) we have the second circle from below of Fig. \ref{fig:3edge_optimal}(see also Fig. \ref{fig:3edge_optimal}.ii.
Indeed, the triangle is tangent to the circle in the medium point $(-\mu,\mu/2,\mu/2)$such that Eq. \eqref{eq:edge_eq} becomes $|a-1|\ge\mu=b=(3-a)/2$ giving $a \le 5/3\simeq 1.66$. 
If $a \in (5/3,2)$ then the shape of positive maps is the intersection of the circle and the triangle (\ref{vert}) and the set of optimal points are three arcs, e.g. the central one of \ref{fig:3edge_optimal} at $a=1.85$. If $a\ge 2$ then there are three optimal points related to the vertices of the triangle.
The altitude $a=2$ coincides with the joining points $[-1,1/2,1/2,2]$, $[1/2,1/2,-1,2]$, and $[1/2,-1,1/2,2]$ of the sea-anchor in Fig. \ref{fig:3edge_optimal}. The red circle inscribes the triangle. If $a \in (2,3]$ then there are three optimal points related to the vertices of the triangle each with one spanning vector, as the circle at the top of Fig. \ref{fig:3edge_optimal} intersects the three black lines (see also Fig. \ref{fig:3edge_optimal}.v) . Notice that the condition from Eqs. (\ref{edge1}-\ref{edge3}) of having the same curve leads to the saturation of the interior condition $a+b+c=3$. This gives seven spanning vectors.
}

\begin{figure}
	    \centering
	    \begin{tikzpicture}
            \node[align=center] at (0,0) {\includegraphics[width=.8\textwidth]{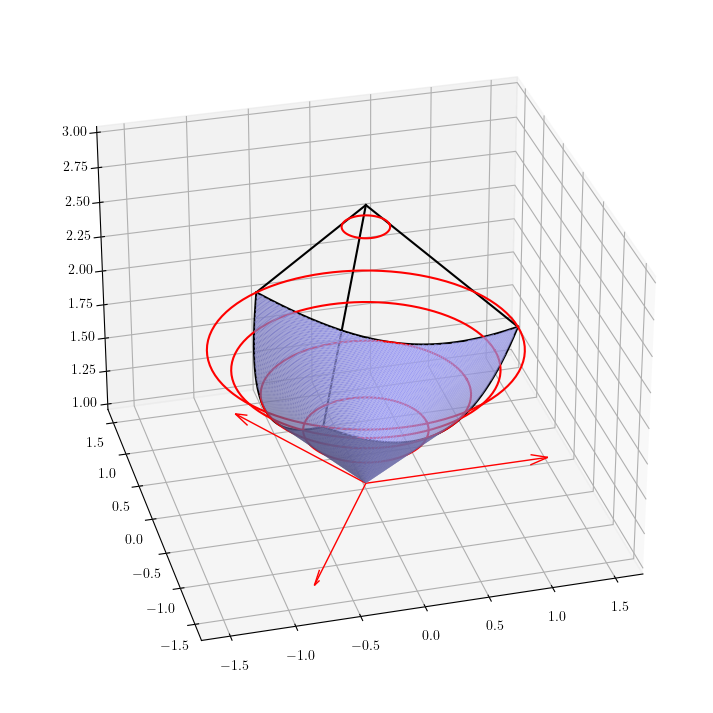}};
            \node[align=center] at (-6.6,1.5) {\large $a$};
            \node[align=center] at (-2.5,-2.2) {\large $d$};
            \node[align=center] at (-.5,-4.5) {\large $e$};
            \node[align=center] at (4,-2.8) {\large $f$};
            \node[align=center,rotate=8] at (-2.5,-4) {$d+e+f=0$ \\ plane};
            \node[align=center] at (0,3.5) {$[0,0,0,3]$};
            \node[align=center] at (-2.8,1.4) {$[-1,\frac 12,\frac 12,2]$};
            \node[align=center] at (4.2,.8) {$[\frac 12,\frac 12,-1,2]$};
            \node[align=center] at (.2,-1.2) {$[\frac 12,-1,\frac 12,2]$};
            \node[align=center] at (-.7,2.8) { (i)};
            \node[align=center] at (-3.2,.7) {(ii)};
             \node[align=center] at (-2.55,.4) {(iii)};
             \node[align=center] at (-1.7,.0) {(iv)};
             \node[align=center] at (-.9,-.9) {(v)};
        \end{tikzpicture} 
     \vspace{5pt}
      \begin{minipage}{0.19\textwidth}
        \includegraphics[width=\linewidth]{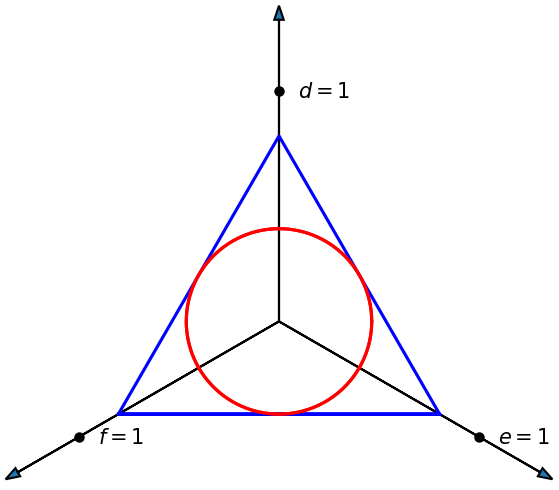}
         \caption*{(i) $a=1.4$}
     \end{minipage}
     \hfill
     \begin{minipage}{0.19\textwidth}
         \includegraphics[width=\linewidth]{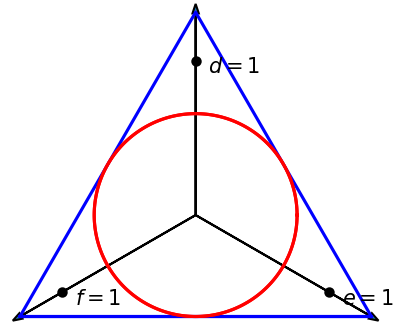}
         \caption*{(ii) $a=\frac 53\simeq 1.66$}
     \end{minipage}
     \hfill
     \begin{minipage}{0.19\textwidth}
         \includegraphics[width=\linewidth]{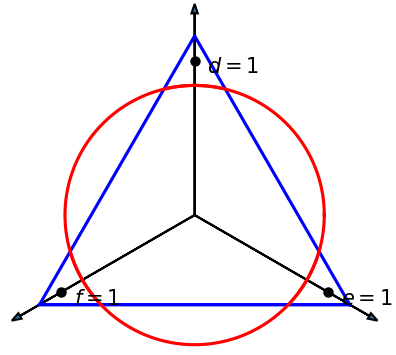}
         \caption*{(iii) $a=1.85$}
     \end{minipage}
     \hfill
     \begin{minipage}{0.19\textwidth}
         \includegraphics[width=\linewidth]{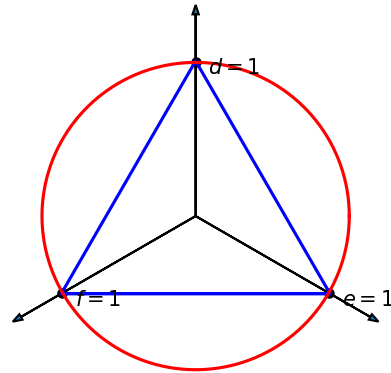}
         \caption*{(iv) $a=2$}
     \end{minipage}
     \hfill
     \begin{minipage}{0.19\textwidth}
         \includegraphics[width=\linewidth]{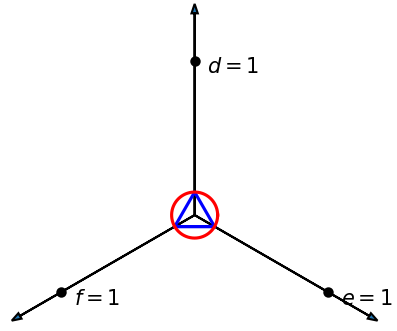} 
         \caption*{(v) $a=2.85$}
     \end{minipage}
     \caption{The set of optimal points when $b=c,~ a+b+c=3$ (when three edge conditions coincide). The sea-anchor shape is the intersection of the cone coming from the edge condition {(red circles)} and the pyramid coming from the vertex condition {by varying the parameter $a$ along the bisector line in Fig. \ref{Kye_ellipse_split_2}}. The coordinates of distinguished points are given in the $[d,e,f,a]$ format. Enumerated from (i) to (v) the red circles from the bottom to the top of the sea-anchor shape correspond to the configurations in the subfigures for each value of $a$.}
	    \label{fig:3edge_optimal}
\end{figure}
It has been left to consider the situations when the saturation of the interior condition $a+b+c=3$ provides the spanning condition.
Since now, we will assume that $a = 3 - b - c$ and the set of admissible points in the $d+e+f=0$ plane is parameterized by two parameters $b,c$, satisfying the conditions $b,c \ge 0, b+c-1 \ge \sqrt{bc}$. The set of admissible pairs $b,c$ is shown in Fig. \ref{Kye_ellipse}.
	
	\begin{figure}
	    \begin{tikzpicture}
	        \begin{scope}[rotate=45, scale = 3]
	          \draw[pattern=north west lines] (2^-.5,2^-.5) -- (0,0) -- (2^-.5,-2^-.5) arc(-120:120:2^.5/3 and 2^.5/3^.5);
	        \end{scope}
	        \draw[->] (0,0) -- (0,4.5);
	        \draw[->] (0,0) -- (4.5,0);
	        \node at (4.2,.2) {$b$};
	        \node at (.2,4.2) {$c$};
	        \fill[black] (0,3) circle (.05);    \node at (-0.6,3) {$(0,1)$};
	        \fill[black] (3,0) circle (.05);    \node at (3,-.4) {$(1,0)$};
	        \fill[black] (3,3) circle (.05);    \node at (3.5,3) {$(1,1)$};
	    \end{tikzpicture}
	    \caption{The set of admissible pairs of parameters, denoted as $b,c$ if the optimality necessary condition $a+b+c=3$ is satisfied.}
	    \label{Kye_ellipse}
	\end{figure}
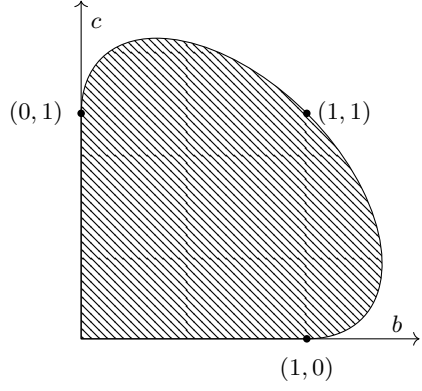

    The typical shape of the set of admissible points in the $d+e+f=0$ plane is now an intersection of 
    {Alice} (edge conditions),  
    {Bob} (vertex conditions) and hence its boundary consists of nine continuous parts (three linear and six curved). 
    {If Alice contains Bob or vice versa,}
    then the smaller {set} 
    is the set of admissible $(d,e,f)$ points. See the figure \ref{fig:triangles} for examples.
    {In the following, we need to combine the above subspaces to span the whole $\mathbb{C}^3\otimes\mathbb{C}^3$. As they have non-trivial intersections, the spanning property is not guaranteed by a simple summing of the dimensions. Fig. \ref{fig:lin} shows in a pictorial representation how to combine properly the spanning vectors from the subspaces generated by the saturation of interior condition (i) with dimension seven, edge condition (e), with dimension 3, and vertex condition (v) with dimension 1.}
  \begin{figure}[h!]
        \centering
        \vspace{5pt}
      \begin{minipage}{0.19\textwidth}
        \includegraphics[width=\linewidth]{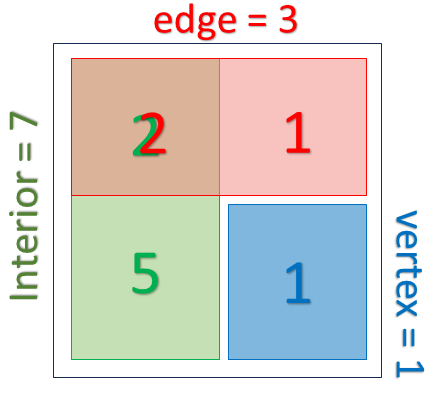}
         \caption*{(iev)}
     \end{minipage}
     \hfill
     \begin{minipage}{0.19\textwidth}
         \includegraphics[width=\linewidth]{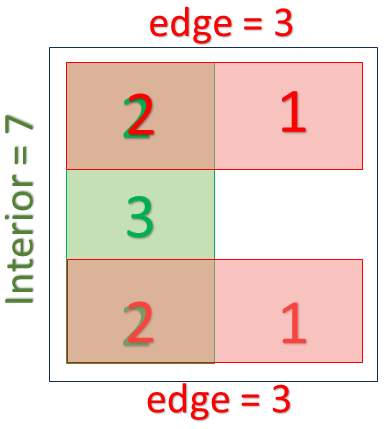}
         \caption*{(iee)}
     \end{minipage}
     \hfill
     \begin{minipage}{0.19\textwidth}
         \includegraphics[width=\linewidth]{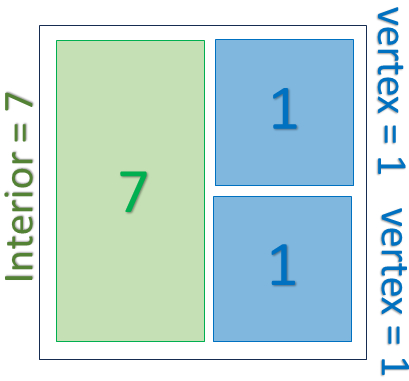}
         \caption*{(ivv)}
     \end{minipage}
     \caption{Representation of the linearly independent vectors to obtain spanning property. Notice that if the interior is satisfied, the spanning vectors contain two vectors coming from the edge condition, that ultimately contributes only with one linearly independent vector. The sum gives 9 linearly independent vectors. Here, the no. of spanning vectors coming from the interior, edge and vertex conditions correspond to the green, red and blues colors, respectively. The overlapping colors (green and red)  indicate the linear dependent vectors. }
     \label{fig:lin}
 \end{figure}

    \begin{figure}[h!]
        \centering
        \begin{minipage}{.32\textwidth}
            \includegraphics[width=\textwidth]{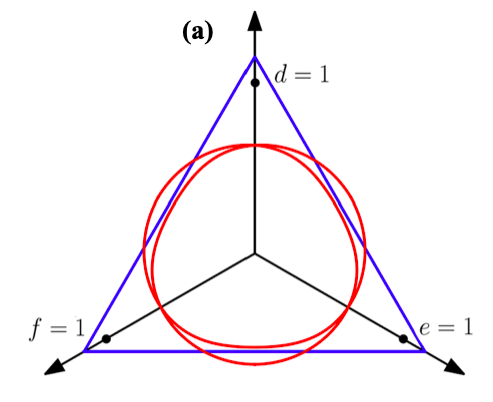} 
        \end{minipage}
        \begin{minipage}{.32\textwidth}
            \includegraphics[width=\textwidth]{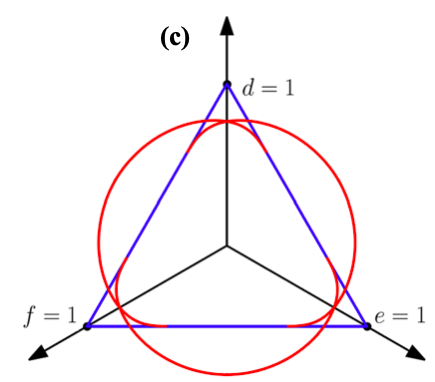}
        \end{minipage}
        \begin{minipage}{.32\textwidth} 
            \includegraphics[width=\textwidth]{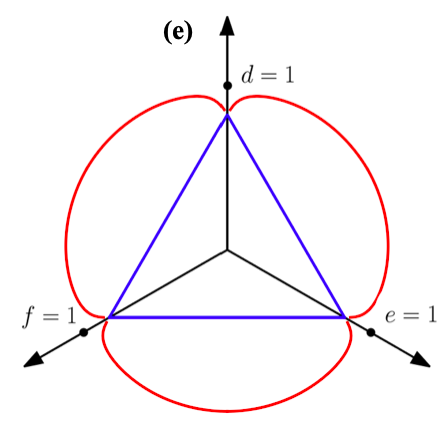}
        \end{minipage}
        \\
        \begin{minipage}{.32\textwidth} 
            \includegraphics[width=\textwidth]{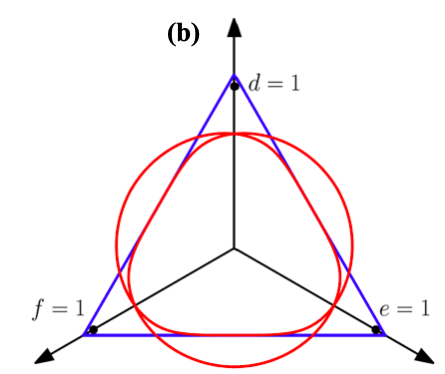}
        \end{minipage}
        \begin{minipage}{.32\textwidth} 
            \includegraphics[width=\textwidth]{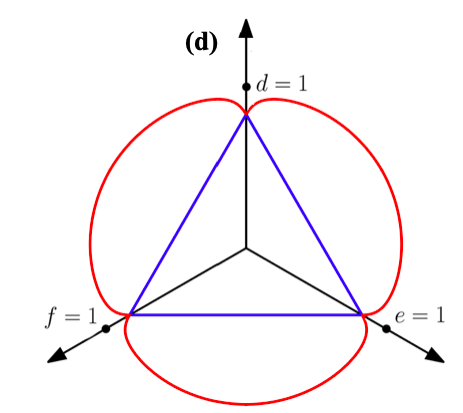}
        \end{minipage}
        \caption{Sets of admissible points in the $d+e+f=0$ plane for $a+b+c=3$, $b-c = 0.2$ and five different choices of $b+c$.  Such a set is an intersection of 
        {Bob} (blue) and 
        {Alice} (red). The row above presents the generic situations when (a) 
        {Alice is contained in Bob,} 
        (c) 
        {Alice and Bob} 
        have a nontrivial intersection, and 
        (e) 
        {Bob is contained in Alice}
        (the values of $b+c$ are $1.35$, $1.20$ and $1.01$ respectively). The row below presents the passages between the above situations: (b) 
        {the boundary of Alice touches the side of Bob}
         at one point ($b+c = 38/30 \approx 1.27$), and (d) arcs of 
         {vertices of Bob coincide with vertices of Alice}
         ($b+c=139/135 \approx 1.03$).} 
        \label{fig:triangles}
    \end{figure}

    \item {\bf Spanning by the saturation of the interior condition with vertex or edge condition:} The saturation of the interior condition provides vectors spanning the subspace $\mathrm{span} \{ \ket{i}\otimes\ket{j}, i\ne j \}$ and one linearly independent vector $[1,1,1]$ in the subspace $\mathrm{span}\{\ket{i}\otimes\ket{i}, i=1,2,3\}$. With these spanning vectors derived from the interior condition, it is depicted in Fig. \ref{fig:lin} that both an edge condition and a vertex condition individually contribute precisely one linearly independent vector to this subspace. Consequently, in addition to satisfying the fully saturated interior condition, one must also satisfy either two edge conditions (iee), an edge condition and a vertex condition (ive), or two vertex conditions (ivv). The first situation is related to a vertex of 
    {Alice}, the second situation is related to the crossing of 
    {Alice and Bob}, and the third situation is a vertex of 
    {Bob}.

    Now we will describe the regions in the set in the Fig. \ref{Kye_ellipse} referring to classes of shapes described in the Fig. \ref{fig:triangles}. First, on the Fig. \ref{Kye_ellipse_split} let us divide the set into subsets depending on the minimum of $a-1,b,c$:
    
    \begin{figure}[h!]
	    \centering
	    \begin{tikzpicture}
	        \begin{scope}[rotate=45, scale = 3]
	          \draw[pattern=north west lines] (2^-.5,2^-.5) -- (0,0) -- (2^-.5,-2^-.5) arc(-120:120:2^.5/3 and 2^.5/3^.5);
	        \end{scope}
	        \draw[->] (0,0) -- (0,4.5);
	        \draw[->] (0,0) -- (4.5,0);
	        \node at (4.2,.2) {$\Large b$};
	        \node at (.2,4.2) {$\Large c$};
	        \fill[black] (0,3) circle (.05);    \node at (-0.6,3) {$(0,1)$};
	        \fill[black] (3,0) circle (.05);    \node at (3.5,-0.4) {$(1,0)$};
	        \fill[black] (3,3) circle (.05);    \node at (3.5,3) {$(1,1)$};
	        \fill[black] (2,2) circle (.05);    \node at (2.5,1.9) {$(\frac 23,\frac 23)$};
	        \draw (2,2) -- (0,3);
	        \draw (2,2) -- (3,0);
	        \draw (0,0) -- (2,2);
	        \node at (1.8,1) {$\mu = c$};
	        \node at (.6,1.5) {$\mu = b$};
	        \node at (2.3,2.6) {$\mu = a-1$};
	    \end{tikzpicture}
	    \caption{Decomposition of the set of admissible pairs $(b,c)$ to subsets depending on the value of $\mu = \min \{a-1,b,c\}$.}
	    \label{Kye_ellipse_split}
	\end{figure}
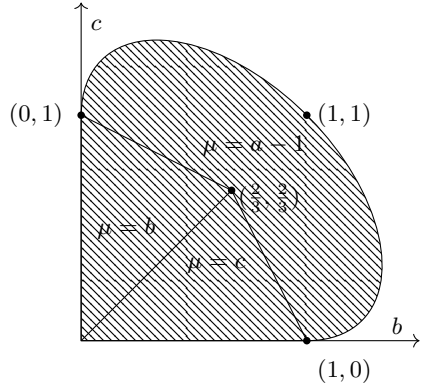

 \begin{figure}[h!]
	    \centering
	    \begin{tikzpicture}
	        \draw[->] (0,0) -- (0,4.5);
	        \draw[->] (0,0) -- (4.5,0);
	        \node at (4.2,.2) {$b$};
	        \node at (.2,4.2) {$c$};
            \begin{axis}[hide axis, xmin=0,   xmax=1.5, ymin=0,   ymax=1.5, yscale=.79, xscale=.66]
                \addplot[name path=F,blue!50,domain={0:.5}] {((4-x)-(9*x^2+4)^.5)/2};
                \addplot[name path=G,blue!50,domain={.5:1}] {((x-2)+((x-2)^2+8*(x^2-4*x+3))^.5)/4};
                \addplot[name path=H,blue,domain={0:1}] {0};
                \addplot[name path=I,blue!50!red,domain={0:2/3}] {1-x/2};
                \addplot[name path=J,blue!50!red,domain={2/3:1}] {2-2*x};
                \addplot[name path=K,red!50,domain={0:4/3},samples=100] {(2-x+2*(x*(1-.75*x))^.5)/2};
                \addplot[name path=L,red!50,domain={1:4/3}] {(2-x-2*(x*(1-.75*x))^.5)/2};
                \addplot[blue!50]fill between[of=F and H, soft clip={domain=0:.5}];
                \addplot[blue!50]fill between[of=G and H, soft clip={domain=.5:1}];
                \addplot[blue!50!red]fill between[of=F and I, soft clip={domain=0:.5}];
                \addplot[blue!50!red]fill between[of=G and I, soft clip={domain=.5:2/3}];
                \addplot[blue!50!red]fill between[of=G and J, soft clip={domain=2/3:1}];
                \addplot[red!50]fill between[of=I and K, soft clip={domain=0:2/3}];
                \addplot[red!50]fill between[of=J and K, soft clip={domain=2/3:1}];
                \addplot[red!50]fill between[of=L and K, soft clip={domain=1:4/3}];
            \end{axis}
            \draw[red] (2,2) -- (3,3);
            \draw[red] (2,2) -- (0,3);
            \draw[red] (2,2) -- (3,0);
            \fill[black] (0,3) circle (.05);    \node at (-0.6,3) {$(0,1)$};
	        \fill[black] (3,0) circle (.05);    \node at (3.5,-0.4) {$(1,0)$};
	        \fill[black] (3,3) circle (.05);    \node at (3.5,3) {$(1,1)$};
	        \fill[black] (2,2) circle (.05);    \node at (2.5,1.9) {$(\frac 23,\frac 23)$};
	        \fill[black] (1.5,1.5) circle (.05);    \node at (1,1.5) {$(\frac 12,\frac 12)$};
	    \end{tikzpicture}
	    \caption{Subset of the set of admissible pairs $(b,c)$ related to different shapes in the plane $d+e+f=0$. 
        {The blue region identifies the blue triangle contained in the red curved-arc triangle, the red region refers to the opposite case, and the mixing of the colors indicates their nontrivial intersection.}}
	    \label{Kye_ellipse_split_2}
	\end{figure}
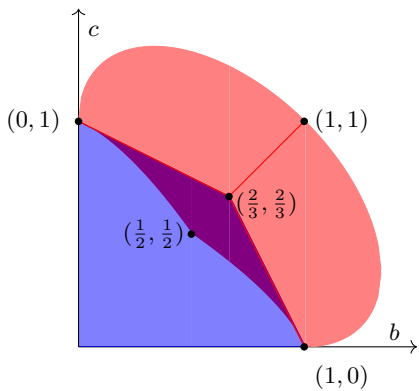
    
    {Alice is contained in Bob} (see Fig.~\ref{fig:triangles} (a)) iff the center of the side $(\frac \mu2 , \frac \mu2, - \mu)$ does not satisfy the edge condition, hence it must be:
    \begin{equation}
        \sqrt{(a+e-1)(a+d-1)} + \sqrt{(b+f)(c+f)} = \sqrt{\left(a-1+\frac \mu2 \right)\left(a-1 + \frac \mu2 \right)} + \sqrt{(b-\mu)(c-\mu)} < 1.
    \end{equation}
    Depending on the value of $\mu$, the above equation takes a form:
    \begin{align}
        b+c-1 >  \frac 12 \min\{ b,c \} & ~~~ \text{if} ~~~ \mu = \min \{ b,c \}, \\
        (b-c)^2 > 0  & ~~~ \text{if} ~~~ \mu = a-1.
    \end{align}
    The first possibility is satisfied only with saturation on line segments connecting $(\frac 23, \frac 23)$ with $(1,0)$ and $(0,1)$. The second inequality is always satisfied if $a-1<\min\{b,c\}$ and saturates on the line segment connecting $(\frac 23, \frac 23)$ with $(1,1)$.
    
    On the other way around, 
    {Bob is contained in Alice}
    (see Fig.~\ref{fig:triangles} (d)) iff the edge $(2\mu,-\mu,-\mu)$ satisfies the edge condition:
    \begin{equation}
        \sqrt{(a+e-1)(a+d-1)} + \sqrt{(b+f)(c+f)} = \sqrt{(a-1+2\mu)(a-1-\mu)} + \sqrt{(b-\mu)(c-\mu)} \ge 1.
    \end{equation}
    Depending on the value of $\mu$, the above equation takes a form:
    \begin{align}
        (a-1+2\mu)(a-1-\mu) \ge 1,  & ~~~ \text{if} ~~~ \mu = \min \{ b,c \}, \\
        (b-a+1)(c-a+1) \ge 1,  & ~~~ \text{if} ~~~  \mu = a-1.
    \end{align}
    The first inequality describes a set below two hyperbolic arcs in Fig.~\ref{Kye_ellipse_split_2}:
    \begin{align}
        3-2b-4c-2b^2+bc+c^2 \ge 0,  & ~~~ \text{if} ~~~ b \le c, \\
        3-2c-4b-2c^2+cb+b^2 \ge 0,  & ~~~ \text{if} ~~~ c \le b,
    \end{align}
    starting from $(0,1)$ and $(1,0)$ respectively and joining when $b=c$.
    The second inequality is satisfied iff $a=b=c=1$ giving the well-known reduction map \cite{reduction1,reduction2}.

    Figure \ref{Kye_ellipse_split_2} shows the regions on the $b,c$ plane, where the corresponding shape on the $d+e+f=0$ plane is: 
    {Alice (the red area), Bob (the blue area)}
    , and the polygon formed by the intersection of 
    {Alice and Bob}
    (the violet area).
    {
     The code in Ref. \cite{code}
    shows the shape of positive maps in the $(d,e,f)$-plane related to the point on the $(b,c)$-plane.}

    For  $(b,c)$-pair in the blue area, one has three optimal points (ivv) being vertices of 
    {Bob}. Their $d,e,f$ coordinates are:
    \begin{equation} 
    \label{opt_ivv}
    (d,e,f) = (2\mu,-\mu,-\mu), \quad \lor \quad (d,e,f) = (-\mu,2\mu,-\mu), \quad \lor \quad (d,e,f) = (-\mu,-\mu,2\mu),
    \end{equation}
    where $\mu = \min\{b,c\}$.

    For  $(b,c)$-pair in a violet region, one has three optimal points being vertices of 
    {Alice}
    and six optimal points (ive) being intersections of 
    {Alice and Bob}. The coordinates of the latter ones are all permutations of the following vector:
    \begin{align} 
    \label{opt_ive}
    \Big[ -\mu, ~
    \frac 12 \left( \mu + \sqrt{\mu^2 - 4(a-1+\mu)(a-1)+4} \right),~
    \frac 12 \left( \mu - \sqrt{\mu^2 - 4(a-1+\mu)(a-1)+4} \right)
    \Big]
    \end{align}

    The remaining three optimal points (iee) being the vertices of 
    {Alice} cannot be expressed by analytical formulas (being roots of a four-degree polynomial). The same aplies to the red region, where each pair $b,c$ provides three optimal points (iee).

    One can show that the optimal maps in Eqs. (\ref{opt_ivv}) and (\ref{opt_ive}) satisfy (saturating) the condition (\ref{hess2}). The same can be shown numerically for optimal maps (iee), hence the Hessian condition becomes idle if the interior condition is saturated.
\end{itemize}

\section{Conclusions}
\label{conclusion}

We provided a complete analysis of optimality for recently proposed positive maps in the matrix algebra $M_3$ \cite{bera22}. It generalizes the analysis for the circulant scenario corresponding to $d=e=f=0$ \cite{KOREA,kye12}. 
In the circulant scenario, these maps are not only optimal but even nd-optimal (or bi-optimal), i.e. the map composed with transposition is still optimal. It would be interesting to analyze this property beyond the circulant scenario. A positive nd-optimal maps cannot be decomposed into positive and decomposable maps and hence provide a powerful tool    
for detecting PPT entangled states \cite{sarbicki14}. Another issue is extremality. In the circulant case \cite{Korea1} it was proved that if $a+b+c=3$, then apart from the reduction map corresponding to $(a,b,c)=(1,1,1)$ all maps are extremal (and apart $(2,1,0)$ and $(2,0,1)$ they are exposed). It would be interesting to provide the corresponding analysis of extremality beyond the circulant case that we have presented. 
    { In the circulant case, the optimal maps are marked green on Figure \ref{fig_kye}, but it can be shown, that each such point is a convex combination of a point from ellipse and a point from the hyperbola, both from the boundary of the region of optimal points in the family. Hence to detect all the entanglement visible by the family, it is enough to use only the maps from the hyperbola and the ellipse. It would be very interesting, from the point of view of efficient detection to find such linear dependencies between optimal elements also in the extended family. {We characterized optimal maps in a domain of parameters contained in a wider space whose parameters indicate entanglement witnesses derived in \cite{sarbicki2020b} which, in turn, extend to the multipartite version in \cite{sarbicki2020a}}.
    }

\section*{Acknowledgements}
This project is supported by QuantERA/2/2020, an ERA-Net co-fund in Quantum Technologies, under the eDICT project. GSc thanks the Institute of Physics of the Nicolaus Copernicus University of Toru\'n for the hospitality and he is also partially supported by Istituto Nazionale di Fisica
Nucleare (INFN) through the project ''QUANTUM''. AB is supported by the Polish National Science Centre through the 
SONATA BIS project No. 2019/34/E/ST2/00369. GSa and DC were supported by the Polish National Science Centre project No. 2018/30/A/ST2/00837.


\appendix
\section{Explicit calculations}
In this appendix, we extend the calculations for the saturation of interior, edge and vertex conditions for a wider readership.
\subsection{Interior condition saturated}
\label{sec:interior}
 The spanning criterion referred to the entanglement witnesses (the Choi matrix $\mathcal{C}(\Phi_W)$) is applied as in the text on the map by the following relation $\bra{\psi^*\otimes \phi}\mathcal{C}(\Phi_W)\ket{\psi^*\otimes \phi}=\bra{\phi}\Phi_W(\ket{\psi}\bra{\psi})\ket{\phi}$. Explicitly $\Phi_W(\ket{\psi}\bra{\psi})$ reads as:
\begin{equation}\label{eq:W_psi}
	\left(\begin{array}{ccc}
		(w_{11}-1)|\psi_1|^2+w_{12}|\psi_2|^2+w_{13}|\psi_3|^2&-\psi_1\psi_2^*&-\psi_1\psi_3^*\\
		-\psi_2\psi_1^*&w_{21}|\psi_1|^2+(w_{22}-1)|\psi_2|^2+w_{23}|\psi_3|^2&-\psi_2\psi_3^*\\
		-\psi_3\psi_1^*&-\psi_3\psi_2^*&w_{31}|\psi_1|^2+w_{32}|\psi_2|^2+(w_{33}-1)|\psi_3|^2
	\end{array}\right).
\end{equation} 
The condition $\det\Phi_W(\ket{\psi}\bra{\psi})=0$ after a straightforward calculation gives
\begin{equation}
	\frac{|\psi_1|^2}{w_{11}|\psi_1|^2+w_{12}|\psi_2|^2+w_{13}|\psi_3|^2}+\frac{|\psi_2|^2}{w_{21}|\psi_1|^2+w_{22}|\psi_2|^2+w_{23}|\psi_3|^2}+\frac{|\psi_3|^2}{w_{31}|\psi_1|^2+w_{32}|\psi_2|^2+w_{33}|\psi_3|^2}\le 1.
\end{equation}
From Yamagami lemma \cite{Yamagami}, being $W/w$ doubly stochastic, there exists only one vector that saturates this condition and corresponds to $\ket{\psi}=[\mathrm{e}^{\mathrm{i}\alpha_1},\mathrm{e}^{\mathrm{i}\alpha_2},\mathrm{e}^{\mathrm{i}\alpha_3}]^T$. Bistochasticity and gauge condition give $a+b+c= 3$. Given $\ket{\psi}$, now the vector $\ket{\phi}$ is defined to be the only element (up to a multiplicative constant) in the kernel of $\Phi_W(\ket{\psi}\bra{\psi})$, i.e. $\Phi_W(\ket{\psi}\bra{\psi})\ket{\phi}=0$,
\begin{align}\label{eq:phiint}
	\left\{
	\begin{array}{l}
		[(w_{11}-1)|\psi_1|^2+w_{12}|\psi_2|^2+w_{13}|\psi_3|^2]\phi_1-\psi_1\psi_2^*\phi_2-\psi_1\psi_3^*\phi_3=0\\
		-\psi_2\psi_1^*\phi_1+[w_{21}|\psi_1|^2+(w_{22}-1)|\psi_2|^2+w_{23}|\psi_3|^2]\phi_2-\psi_2\psi_3^*\phi_3=0\\
		-\psi_3\psi_1^*\phi_1-\psi_3\psi_2^*\phi_2+[w_{31}|\psi_1|^2+w_{32}|\psi_2|^2+(w_{33}-1)|\psi_3|^2]\phi_3=0
	\end{array}
	\right.
\end{align}
which is satisfied for $\ket{\phi}=\ket{\psi}=[\mathrm{e}^{\mathrm{i}\alpha_1},\mathrm{e}^{\mathrm{i}\alpha_2},\mathrm{e}^{\mathrm{i}\alpha_3}]^T$, (i.e. $x=\bm 1=[1,1,1]$ in the main text since $|\psi_i|^2=|\phi_i|^2=1$). We highlight here that $\ket{\phi}$ is the only vector in the kernel (the cardinality of the kernel corresponds to the number of zero eigenvalues) because by definition the positive semidefinite $\Phi_W(\ket{\psi}\bra{\psi})$ can have only one eigenvalue equal to zero along the direction $\phi=\psi$ as $\Phi_W(\ket{\psi}\bra{\psi})=\Phi_D(\ket{\psi}\bra{\psi})-\ket{\psi}\bra{\psi}$ with
\begin{equation}
	\bra{\phi}\Phi_D(\ket{\psi}\bra{\psi})\ket{\phi}=
	\sum_{i}w_{ii}|\phi_i|^2|\psi_i|^2+\sum_{i\neq j}w_{ij}|\phi_i|^2|\psi_j|^2\ge
	\sum_{i}|\phi_i|^2|\psi_i|^2.
\end{equation}
For the rank-nullity theorem $\dim \ker \Phi_W(\ket{\psi}\bra{\psi})+\dim \mathrm{rank}\Phi_W(\ket{\psi}\bra{\psi})=3$ there are two vectors orthogonal to $[1,1,1]$ which are $[1,-1,0],[0,1,-1]$. Therefore the saturation of the interior conditions gives 7 spanning vectors for each $\ket{\psi\otimes\phi^*}$ of the kind
$
	[1,\mathrm{e}^{\mathrm{i}\alpha_{12}},\mathrm{e}^{\mathrm{i}\alpha_{13}},\mathrm{e}^{\mathrm{i}\alpha_{21}},1,\mathrm{e}^{\mathrm{i}\alpha_{23}},\mathrm{e}^{-\mathrm{i}\alpha_{13}},\mathrm{e}^{-\mathrm{i}\alpha_{23}},1]^T$ in the orthogonal complement $
[e_1 \otimes e_1- e_2 \otimes e_2, e_2 \otimes e_2 - e_3 \otimes e_3]^\perp$. Alternatively the seven spanning vectors can be count following the appendix \ref{sec:randomphases}.

\subsection{Edge condition saturated}
\label{sec:edgecondition}
In the case of $x_1=0$ and $x_2,x_3\neq 0$ $\Phi_W(\ket{\psi}\bra{\psi})$ looks
\begin{equation}\label{eq:W_psi}
	\left(\begin{array}{ccc}
		w_{12}|\psi_2|^2+w_{13}|\psi_3|^2&0&0\\
		0&(w_{22}-1)|\psi_2|^2+w_{23}|\psi_3|^2&-\psi_2\psi_3^*\\
		0&-\psi_3\psi_2^*&w_{32}|\psi_2|^2+(w_{33}-1)|\psi_3|^2
	\end{array}\right).
\end{equation} 
The edge condition is 
\begin{equation}
	\frac{x_2}{w_{22}x_2+w_{23}x_3}+\frac{x_3}{w_{32}x_2+w_{33}x_3}\ge1
\end{equation}
By multiplying the m.c.m. we obtain
\begin{equation}\label{eq:qf1}
	w_{32}(w_{22}-1)x_2^2+w_{23}(w_{33}-1)x_3^2+[(w_{32}w_{23}+w_{22}w_{33}-w_{22}-w_{33})]x_2x_3\ge0,
\end{equation}
which is a quadratic form of the kind $\alpha x_2+\beta x_3+\gamma x_2x_3\ge0$, with the coefficients $\alpha,\beta,\gamma$.
Notice that $\gamma=[(w_{32}w_{23}+w_{22}w_{33}-w_{22}-w_{33})]=(w_22-1)(w_33-1)+w_{23}w_{32}-1$. Moreover $(\sqrt{\alpha}x_2-\sqrt{\beta}x_3)^2=\alpha x_2^2+\beta x_3^2-2\alpha \beta x_2x_3\ge0$, therefore the edge condition is saturated for $\gamma=-2\alpha \beta$\footnote{The edge condition written in the main text comes from here indeed this expression is a perfect square: $(w_22-1)(w_33-1)+w_{23}w_{32}-1=-2\sqrt{w_{23}w_{32}(w_{22}-1)(w_{33}-1)}$ so that $\sqrt{(w_{22}-1)(w_{33}-1)}+\sqrt{w_{23}w_{32}}\ge1$.}. This means that Eq. \eqref{eq:qf1} can be rewritten as
\begin{equation}
x^TM_{1} x=
	\left[\begin{array}{cc}
		x_2&x_3
	\end{array}\right]
	\left(\begin{array}{cc}
		w_{32}(w_{22}-1)&-\sqrt{w_{32}w_{23}(w_{22}-1)(w_{33}-1)}\\
		-\sqrt{w_{32}w_{23}(w_{22}-1)(w_{33}-1)}&w_{23}(w_{33}-1)\\
	\end{array}\right)
\left[\begin{array}{c}
	x_2\\x_3
\end{array}\right]
\end{equation}
where $\det M_1=0$. This means that its kernel is non-trivial and admits one vector obtained from $M_1 x=0$ which is
\begin{equation}
	w_{32}(w_{22}-1)x_2-\sqrt{w_{32}w_{23}(w_{22}-1)(w_{33}-1)}x_3=0.
\end{equation}
By dividing $\sqrt{w_{32}(w_{22}-1)}$ it ends to
\begin{equation}
	\sqrt{w_{32}(w_{22}-1)}x_2-\sqrt{w_{23}(w_{33}-1)}x_3=0.
\end{equation}
Therefore, up to a multiplicative constant, we have
\begin{equation}
	x_2=[w_{32}(w_{22}-1)]^{-\frac12}=[(c+f)(a+e-1)]^{-\frac12},\qquad
	x_3=[2_{23}(w_{33}-1)]^{-\frac12}=[(b+f)(a+d-1)]^{-\frac12}.
\end{equation}
The vector $\ket{\psi}$ is immediately obtained in Eq. \eqref{eq:edgepsi}.
Now the vector $\ket{\phi}$ is found in the kernel of $\Phi_W(\ket{\psi}\bra{\psi})$ which is
\begin{align}\label{eq:phiedge}
	\left\{
	\begin{array}{l}
		(w_{12}x_2+w_{13}x_3)\phi_1 = 0,\\
		((w_{22}-1)x_2+w_{23}x_3)\phi_2 - \sqrt{x_1x_2}e^{\mathrm{i}(\alpha_2-\alpha_3)}\phi_3 = 0,\\
		-\sqrt{x_1x_2}e^{-\mathrm{i}(\alpha_2-\alpha_3)}\phi_2 + [w_{32}x_2+(w_{33}-1)x_3]\phi_3 = 0
	\end{array}
	\right.
\end{align}
with solution (up to a multiplicative constant)
\begin{equation}
	\ket{\phi}=[0,\sqrt{x_2x_3}e^{\mathrm{i}\alpha_2},((w_{22}-1)x_2+w_{23}x_3)e^{\mathrm{i}\alpha_3}]
	\propto
	\left[0,\frac{\mathrm{e}^{\mathrm{i}\alpha_2}}{[w_{23}w_{32}(w_{22}-1)(w_{33}-1)]^{1/4}},\left(\sqrt{\frac{w_{22}-1}{w_{32}}}+\sqrt{\frac{w_{23}}{w_{33}-1}}\right)\mathrm{e}^{\mathrm{i}\alpha_3}\right]
\end{equation}
Multiplying by $[w_{32}(w_{33}-1)]^{1/4}$ and using the edge condition $\sqrt{(w_{22}-1)(w_{33}-1)}+\sqrt{w_{23}w_{32}}=1$ we finally obtain a more symmetric expression
\begin{equation}
	\ket{\phi}=\left[0,\frac{\mathrm{e}^{\mathrm{i}\alpha_2}}{[w_{23}(w_{22}-1)]^{1/4}},\frac{\mathrm{e}^{\mathrm{i}\alpha_3}}{[w_{32}(w_{33}-1)]^{1/4}}\right]
	=
	\left[0,\frac{\mathrm{e}^{\mathrm{i}\alpha_2}}{[(b+f)(a+e-1)]^{1/4}},\frac{\mathrm{e}^{\mathrm{i}\alpha_3}}{[(c+f)(a+d-1)]^{1/4}}\right]
\end{equation}
This leads to three spanning vectors showed in Eq. \eqref{eq:edgespan}, which are of the form $[k,s\mathrm{e}^{\mathrm{i}\eta},\frac{\mathrm{e}^{-\mathrm{i}\eta}}{s},\frac1k]$ with $k,s$ constants, thus there are three linear independent vectors for $\eta=0,\pi/2,\pi$ {(see the method of random phases in the next appendix)}. Analogously calculations come from the other two edge conditions.

\subsection{Vertex condition}
\label{sec:vertexcondition}
When two component of $\ket{\psi}$ are vanishing, the the map $\Phi_W(\ket{\psi}\bra{\psi})$ is diagonal and admits nontrivial kernel only if one of the diagonal component is zero which is possible when $w_{ii}=1$, that is the vertex condition saturation. Following the main text, for $x_1=x_2=0$, then the nontrivial kernel is obtained for the vertex condition $w_{33}-1=a+d-1=0$, and it is immediate to have as element in its kernel $\ket{\psi}=e_3$. To find $\ket{\phi}$ as for Eqs. (\ref{eq:phiedge}-\ref{eq:phiint}) we have
\begin{align}\label{eq:phiint}
	\left\{
	\begin{array}{l}
		w_{13}x_3\phi_1=0\\w_{23}x_3\phi_2=0\\(w_{11}-1)x_3\phi_3=0\\
	\end{array}
	\right.
\end{align}
where the first two equation are satisfied for $\phi_1=\phi_2=0$ and the last one by the vertex condition $w_{33}=1$ such that $\ket{\phi}=e_3$ and $\ket{\psi\otimes\phi^*}$ span a one--dimensional vector space.

\section{Saturation in the (eee) scenario}

\subsection{Expression for the gradients}
\label{gradient_expression}
In this section, we obtain the Eqs. (\ref{edgeF1}-\ref{edgeF3}). We are interested only in the direction of the gradient, therefore we have the freedom to multiply the vector by a factor. Therefore for $\vec{\nabla}F=[\partial_d F,\partial_e F,\partial_f F]\Big|_{F=1,~d+e+f=0}$
\begin{align}
	g_1\equiv2\sqrt{(a+f-1)(a+e-1)}\vec{\nabla}F_1 =
	\left[\frac{b+c+2 d}{\sqrt{(b+d) (c+d)}}-(b+c+2 d),a+f-1,a+e-1\right]\\
	g_2\equiv2\sqrt{(a+d-1)(a+f-1)}\vec{\nabla}F_2= 
	\left[a+f-1,\frac{b+c+2 e}{\sqrt{(b+e) (c+e)}}-(b+c+2 e),a+d-1\right]\\
	g_3\equiv2\sqrt{(a+e-1)(a+d-1)}\vec{\nabla}F_3=
	\left[a+e-1,a+d-1,\frac{b+c+2 f}{\sqrt{(b+f) (c+f)}}-(b+c+2 f)\right].
\end{align} 
Now, we remove the component orthogonal to the plane characterized by $d+e+f=0$ by $\Pi_{\bm{1}^\perp}=\mathrm{id}-\ket{\bm{1}}\bra{\bm{1}}$, adjusting the terms using $d+e+f=0$, and then using the saturation conditions:
\begin{align}
	\sqrt{(a+e-1)(a+f-1)}=& 1-\sqrt{(b+d) (c+d)},\\
	\sqrt{(a+f-1)(a+d-1)}=& 1-\sqrt{(b+e) (c+e)},\\
	\sqrt{(a+d-1)(a+e-1)}=& 1-\sqrt{(b+f) (c+f)}
\end{align}
we obtain $\Pi_{\bm{1}^\perp}g_i$ for $i=1,2,3$ corresponds to Eqs. (\ref{gradients}) of the main text. For a fast and double check for symbolic calculations, we add a Mathematica file in the supplement materials in the same folder of Ref. \cite{code}. 

\subsection{Derivation of Eq. \eqref{eq:edge_eq}  }
\label{derivation}
From Eq. \ref{equal_vec} we observe that on the LHS $a,b,c$ are constant, therefore we other terms must be considered also constant, namely
\begin{equation}
	\kappa= \frac{b+c+2d}{\sqrt{(b+d)(c+d)}}
	= \frac{b+c+2e}{\sqrt{(b+e)(c+e)}}
	= \frac{b+c+2f}{\sqrt{(b+f)(c+f)}}
\end{equation}
for some constant $\kappa$, which is of the form
\begin{equation}
	\kappa=\frac{t_1^2+t_2^2}{t_1t_2}= \frac{t_1}{t_2}+\frac{t_2}{t_1}.
\end{equation}
Therefore $\kappa$ is a constant iff $t_1/t_2$ is constant. Thus, computing the derivative of $t_1/t_2$ w.r.t. the variables $d,e,f$ that appear independently for each equation we found that $b=c$, and $\kappa=2$, which leads to $a+b+c=3$. Given that, using $a+2b=3$ and $d=-e-f$ Eq. \ref{edge1} becomes $\sqrt{(a+e-1)(a+f-1)}\ge(1-b)-d=(a-1)/2+(e+f)$. Finally, by squaring both the terms 
\begin{equation}
	(a-1)^2+(a-1)(e+f)+ef\ge \frac{(a-1)^2}{4}+(a-1)(e+f)+(e+f)^2
\end{equation}
yields to $3(a-1)^2/4\ge (e+f)^2-ef$, that leads to Eq. \eqref{eq:edge_eq} of the main text by writing $ef=((e+f)^2-e^2-f^2)/2=(d^2-e^2-f^2)/2$. Finally Eq. (\ref{hessian1}-\ref{hessian2}) are obtained for both the signs of the module of hessian condition for $b=c$.

\section{Method of random phase}\label{sec:randomphases}
To satisfy $\left\langle \psi_{i}\otimes\psi_{i}^{*}\right|W\left|\psi_{i}\otimes\psi_{i}^{*}\right\rangle =0$
 the most general vector is $\ket{\psi_i\otimes\psi_i^*}$ with $\ket{\psi_{i}}=[\pm |\psi_{0}^{\left(i\right)}|\mathrm{e}^{i\vartheta_{0}^{\left(i\right)}},\pm|\psi_{1}^{\left(i\right)}|\mathrm{e}^{i\vartheta_{1}^{\left(i\right)}}\pm|\psi_{2}^{\left(i\right)}|\mathrm{e}^{i\vartheta_{2}^{\left(i\right)}}]^T$. There exist 7 linearly independent vectors, for $i=1,\dots 7$, fixing $[|\psi_{0}^{\left(i\right)}|,|\psi_{1}^{\left(i\right)}|,|\psi_{2}^{\left(i\right)}|]$ and varying the phase such that \begin{equation}
\sum_{j=0}^{2}\frac{|\psi_{j}|^2}{\sum_{i=0}^{2}|\psi_{i}|^2w_{ji}+|\psi_{j}|^{2}}=1.
\end{equation} 
Indeed, 
\begin{equation}
\left(\begin{array}{c}
\psi_{1}\mathrm{e}^{\mathrm{i}\theta_{1}}\\
\vdots\\
\psi_{n}\mathrm{e}^{\mathrm{i}\theta_{n}}
\end{array}\right)\otimes\left(\begin{array}{c}
\psi_{1}^{*}\mathrm{e}^{-\mathrm{i}\theta_{1}}\\
\vdots\\
\psi_{n}^{*}\mathrm{e}^{-\mathrm{i}\theta_{n}}
\end{array}\right)=\left(\begin{array}{cccccccccc}
\left|\psi_{1}\right|^{2} & \cdots & \psi_{1}\psi_{n}^{*}\mathrm{e}^{\mathrm{i}\theta_{1n}} & \psi_{2}\psi_{1}^{*}\mathrm{e}^{\mathrm{i}\theta_{21}} & \left|\psi_{2}\right|^{2} & \cdots & \psi_{2}\psi_{n}^{*}\mathrm{e}^{\mathrm{i}\theta_{2n}} & \cdots & \psi_{n}\psi_{1}^{*}\mathrm{e}^{\mathrm{i}\theta_{n1}} & \dots\end{array}\left|\psi_{n}\right|^{2}\right)^{T}
\end{equation}
with $\theta_{kl}=$$\theta_{k}-\theta_{l}$ there are $n^{2}-n+1$
linearly independent vectors by varying $\theta_{kl}.$ Given
$\vec{\theta}_{1}=\left(\theta_{1},\dots\theta_{n}\right)$ we generate
$\left(\theta_{12},\dots,\theta_{1n};\theta_{23},\dots,\theta_{2n};\dots,\theta_{n-1,n}\right)$
linearly independent directions from $0$ to $\pi$, and the opposite
directions linearly independent are obtained swapping the indices,
therefore there are $2\frac{n\left(n-1\right)}{2}$ different directions.
The proof is concluded by adding the direction pointing to $1$ in the unit complex circle. A construction is available in the GitHub folder of Ref. \cite{code}.
\end{document}